\documentclass{pas}

\usepackage{multirow}

\usepackage{graphicx}

\begin{document}

\lefttitle{Publications of the Astronomical Society of Australia}
\righttitle{Xiangcun Meng}

\jnlPage{1}{4}
\jnlDoiYr{2021}
\doival{10.1017/pasa.xxxx.xx}

\articletitt{Research Paper}

\title{Eccentric millisecond pulsar + subdwarf B star from rotationally delayed accretion-induced-collapse scenario}

\author{\sn{Meng} \gn{Xiangcun}$^{1}$}

\affil{$^1$International Centre of Supernovae (ICESUN), Yunnan Key Laboratory of Supernova Research, 
Yunnan Observatories, Chinese Academy of Sciences (CAS), Kunming 650216, China}

\corresp{X. Meng, Email: xiangcunmeng@ynao.ac.cn}

\citeauth{Meng X.,  {\it Publications of the Astronomical Society of Australia} {\bf 00}, 1--12. https://doi.org/10.1017/pasa.xxxx.xx}

\history{(Received xx xx xxxx; revised xx xx xxxx; accepted xx xx xxxx)}

\begin{abstract}
Eccentric millisecond pulsar + helium white dwarf  (MSP + He WD) systems have attracted increasing attention, with the rotationally delayed accretion-induced collapse (RD-AIC) scenario proposed as a possible formation channel. Given the similarity between the formation channels of He WDs and subdwarf B (sdB) stars, eccentric MSP + sdB binaries could also exist in the Galaxy, though none have been detected so far. Theoretical predictions of their properties would greatly aid in their discovery. Here, within the RD-AIC framework, I present predictions for their orbital parameters, including MSP mass, secondary mass, eccentricity and orbital period. Based on two detailed binary population synthesis calculations, I estimate their Galactic birth rate to be $(0.67-1.5)\times10^{\rm -4}~{\rm yr^{\rm -1}}$. Then, a very conservative upper limit for their total number in the Galaxy is 15000, implying that the most optimistic fraction of eccentric MSP + sdB systems among all MSP + sdB populations could reach up to 55\%. These systems are relatively young, with ages on the order of a few hundred Myr, and should therefore be found in relatively young environments. Furthermore, most MSPs in such eccentric binaries have masses below 1.5 $M_{\odot}$. I also briefly discuss their potential future applications in various astrophysical context.
\end{abstract}


\maketitle

\section{INTRODUCTION} \label{sect:1}
Millisecond pulsars (MSPs) are characterized by weak surface
magnetic fields of $B\sim10^{\rm 8}$ G, spin periods shorter
than 30 ms, period derivatives less than $10^{\rm -19}{\rm s\,
s^{\rm -1}}$ and nearly circular orbits. They play a vital role in constraining the equation of state (EoS) of neutron stars (NSs,
\citealt{OZEL16}). The prevailing view is that MSPs form in low-mass X-ray binaries (LMXBs) through a recycling process:
a pulsar that has crossed the death line is spun up to millisecond periods via accretion from a Roche lobe–filling companion. (\citealt{ALPAR82};
\citealt{RADHAKRISHNAN82}; \citealt{BHATTACHARYA91}). Tidal torques during this phase circularize the binary orbit (\citealt{PHINNEY94}). 
This scenario is supported by observed links between LMXBs and MSPs
(\citealt{WIJNANDS98}; \citealt{ARCHIBALD09}),  as well as the fact that 
nearly all fully recycled MSPs with helium white dwarf (WD) companions in the Galaxy exhibit very low orbital eccentricities
(\citealt{MANCHESTER05}).

However, the detection of several eccentric MSPs suggests that the recycling process may be more complex than previously envisioned. 
These eccentric MSPs are found with either main-sequence (MS) stars or low-mass helium WDs as companions
(\citealt{CHAMPION08}; \citealt{BARR17}; \citealt{OCTAU18};
\citealt{STOVALL19}; \citealt{ZHUWW19}; \citealt{SERYLAK22}). While several formation mechanisms have been proposed, 
the origin of such systems remains uncertain. Most models invoke either asymmetric mass loss or environmental interactions: 
I) Dynamical perturbations in a hierarchical triple system may eject one component, leaving an eccentric binary (\citealt{FREIRE11}). 
II) A circumbinary (CB) disk formed from mass loss in the progenitor system may interact with the inner binary, exciting orbital eccentricity
(\citealt{ANTONIADIS14}). 
III) Asymmetric mass ejection during proto-WD formation could impart a kick, resulting in an eccentric MSP binary
 (\citealt{HANQ21}; \citealt{TANGWS23}). 
 IV) The accretion-induced collapse (AIC) of a rapidly rotating super-Chandrasekhar WD could directly produce an MSP, with mass loss inducing eccentricity (\citealt{FREIRE14}; \citealt{WANGD23}). 
 V) Similarly, a phase transition from a rapidly rotating NS to a strange star, accompanied by sudden mass loss, may also yield an eccentric MSP binary (\citealt{JIANGL15}). VI) Resonant interaction with convective flows in a tidally locked red giant, when the eddy turnover time matches the spin period, could drive large eccentricities
(\citealt{GINZBURG21}). 

Hot subdwarfs, identified as extreme horizontal branch (EHB) stars in the Hertzsprung–Russell (HR) diagram, are helium-core-burning objects with very thin hydrogen envelopes. (\citealt{HEBER09,HEBER16}). Like helium WDs, they typically form from red giants that have lost their hydrogen envelopes. Hot subdwarfs are important in several astrophysical contexts, for instance, as contributors to the ultraviolet upturn in elliptical galaxies and as potential surviving companions of Type Ia supernovae (\citealt{HANZW07}; \citealt{GEIER15}; \citealt{MENGXC21}).  
After central helium exhaustion,  hot subdwarfs evolve into carbon-oxygen (CO) WDs, including very low-mass WDs with CO cores (\citealt{PRADA09}; \citealt{JUSTHAM11b}).
Spectroscopically, they are classified into several subtypes, e.g. the dominant subgroup, subdwarf B (sdB) stars, exhibits strong hydrogen Balmer lines and weak or absent He I lines (\citealt{MOEHLER90}; \citealt{GEIER17}; \citealt{HERJ25}).

Although neutron stars can, in principle, have diverse companions, no NS binaries with helium-star or hot subdwarf companions had been identified until 2025. Recently, however, NS + hot subdwarf systems have attracted growing theoretical and observational interest. They are potential gravitational wave sources detectable by the Laser Interferometer Space Antenna (LISA), and may represent progenitors of certain peculiar MSP binaries, such as those with CO WD companions or black widows with extremely low-mass companion (\citealt{TAURIS12}; \citealt{WUY18}; \citealt{GOTBERG20}; \citealt{GUOYL24}). For over a decade, several surveys have sought such binaries to understand their apparent rarity and evolutionary pathways (e.g. \citealt{GEIER11}; \citealt{WUY20}). 

The recent report of a compact binary comprising a MSP and a stripped helium star, the first of its kind, has reinvigorated this field (\citealt{YANGYZL25}). This system, consistent with standard recycling predictions, exhibits a circular orbit (\citealt{GUOYL25}). Yet, given that eccentric MSP + He WD systems, though rare, challenge the standard recycling picture, new questions have emerged: Do eccentric MSP + hot subdwarf systems also exist, in contradiction to standard recycling predictions? If so, how many such systems reside in the Galaxy, what are their characteristic properties, and in what kind of environments within the Milky Way can they be discovered? In this paper, I address these questions by predicting the properties of potential eccentric MSP + hot subdwarf binaries. Such predictions will facilitate future searches for these systems, whose discovery could provide strong constraints on the standard recycling theory.

In studies of the properties of the surviving companions of
Type Ia supernovae (SNe Ia) in supernova remnants (SNRs), 
the so-called spin-up/spin-down model suggests that the companion of 
a rapidly rotating CO WD prior to explosion could be an sdB star (\citealt{JUSTHAM11};
\citealt{DISTEFANO12}; \citealt{MENGXC19a}).
If the WD is composed of
oxygen-neon-magnesium (ONeMg), AIC may occur instead of a SN Ia. While direct evidence for AIC events remains elusive, 
indirect observational clues support this channel, e.g. the young pulsars in globulars clusters (\citealt{TAURIS13}; \citealt{WANGB20}; \citealt{KREMER24}). As discussed in Section~\ref{sect:4.1}, the rotationally delayed AIC (RD-AIC) scenario proposed by \citet{FREIRE14}  offers a promising explanation for known eccentric MSP + He WD systems.
If the RD-AIC scenario is valid, the outcome of an AIC event in a rapidly rotating ONeMg WD + sdB binary would be an eccentric MSP + sdB system, or at least an eccentric NS + sdB binary. This work aims to characterize such yet-undetected eccentric MSP + sdB systems and provide observable predictions to guide future searches  (\citealt{OOSTRUM20}).

In section~\ref{sect:2}, I simply describe my methods, and I
present the results in section~\ref{sect:3}. In section~\ref{sect:4}, I show discussions and my main conclusions.

\begin{figure}
\centerline{\includegraphics[angle=270,scale=.40]{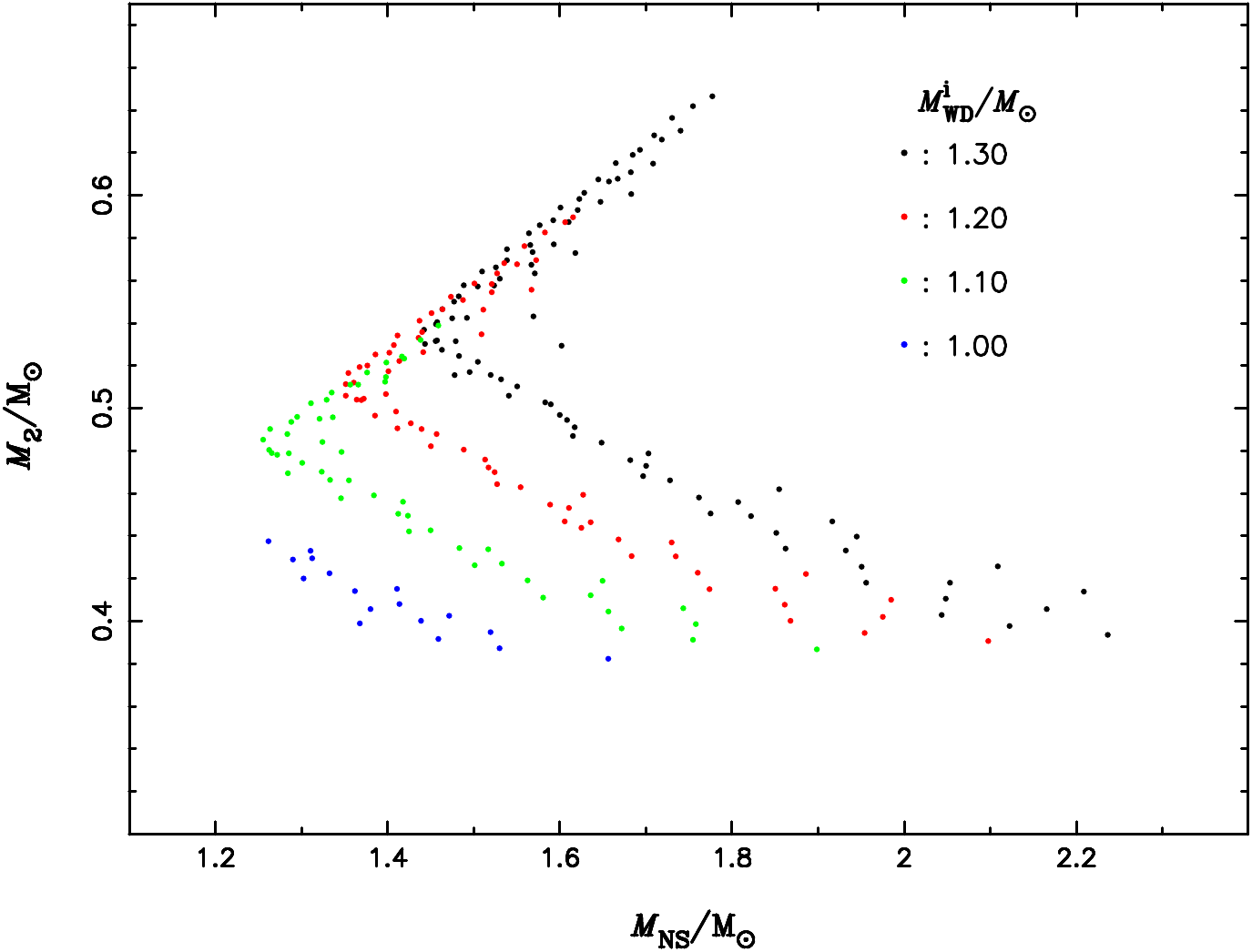}}
\caption{The masses of two components of the eccentric MSP + sdB
systems for different initial WD masses.}\label{mnsm2}
\end{figure}

\section{METHODS}\label{sect:2}
As far as the potential eccentric MSP + sdB systems from the
RD-AIC scenario, their lifetimes may be much shorter than normal
MSPs from recycled scenario because massive WDs tend to have
higher magnetic fields ($10^{\rm 6}\sim10^{\rm 9}$ G,
\citealt{WF05}), which leads to the post-AIC MSPs with a strong
magnetic field of $10^{\rm 11}\sim10^{\rm 14}$ G, as high as
magnetar. Only those from the WDs with very weak magnetic fields
may show the properties of normal MSPs (see also
\citealt{IVANOVA08}). Here, I simply assume that a post-AIC NS
behaves as a MSP no matter how strong its magnetic field is, and
then how long its lifetime. I will discuss this further in
Section~\ref{sect:4}.

Following \citet{MENGXC19a}, the initial binary systems are ONeMg WD +
MS systems, and mass transfer begins when the companions are on
the MS or in the Hertzsprung gap (HG) (see also details in
\citealt{MENGXC17}). The basic methods and assumptions for the
binary evolution are the same to \citet{MENGXC19a} before AIC, and
then I do not give the repetitious details here. As far as the maximum WD is concerned,  a rigid rotation is assumed for the pre-AIC WD and then the maximum WD mass is 1.48 $M_{\odot}$ in \citet{FREIRE14} (see \citealt{MENGXC19a} and
\citealt{TAURIS13}).  Actually, if a differential rotation is considered, the maximum  mass may be up to 4.0 $M_{\odot}$ (\citealt{YOON04,YOON05}). In this paper, I assume that the pre-AIC WD may be differentially rotating if its mass exceeds 1.48
$M_{\odot}$. The WD mass in this paper is then as massive as 2.6 $M_{\odot}$. The rapidly rotating super-Chandrasekhar WDs will experience a spin-down phase before an AIC process. However, at present, many uncertainties are still not resolved in the
spin-up/spin-down model, e.g. the exact spin-down timescale and the exact time of the onset of the spin-down phase (\citealt{MENGXC13}). Following \citet{MENGXC19a} and \citet{MENGXC21}, I define a pseudo-spin-down timescale, $\tau_{\rm sp}=10^{\rm 7}$ yr, in which $\tau_{\rm sp}$ represents a time interval from the moment of $M_{\rm WD}=1.378$ $M_{\odot}$ to the time of AIC (see also \citealt{MENGXC13})\footnote{Here, the real spin-down timescale is generally longer than $9\times10^{\rm 6}$ yrs, since the mass transfer timescale is generally shorter than $10^{\rm 6}$ yrs.}.  For the AIC process, I take similar
assumptions to those in \citet{FREIRE14}. i) The binding energy of
a NS is calculated based on \citet{LATTIMER89} and the baryonic
material of 0.02 $M_{\odot}$ is ejected during AIC
(\citealt{WANGB20}). ii) I assume a circular orbit for the
pre-AIC binary. If the AIC is symmetric, I use
 \begin{equation}
 e=\frac{\Delta M}{M_{\rm NS}+M_{\rm 2}},\label{eq:1}
  \end{equation}
to calculate the eccentricity of the post-AIC system, where
$M_{\rm 2}$, $M_{\rm NS}$ and $\Delta M$ are the pre-AIC secondary
mass, the post-AIC MSP mass and the mass loss during AIC,
respectively (\citealt{BHATTACHARYA91}). This means that the
effect of the ejected baryonic material on the companion is
neglected because of its small amount. I calculate the post-AIC
semimajor axis, $a$, by
 \begin{equation}
 \frac{a}{a_{\rm 0}}=\frac{M_{\rm 0}-\Delta M}{M_{\rm 0}-2\Delta M},\label{eq:2}
  \end{equation}
and then calculate the orbital period of the post-AIC systems
according to $a$, where $a_{\rm 0}$ and $M_{\rm 0}$ are the
orbital separation and the total mass of the pre-AIC system
(\citealt{HILLS83}). iii) Following the formulae in
\citet{HILLS83}, I also discuss the dynamical effect of
asymmetrical AIC on the post-AIC systems by a Monte Carlo way (see
detailed discussions in \citealt{FREIRE14}).

Based on the calculations in \citet{MENGXC19a}, I may get the
parameter space in the initial orbital period - secondary mass
plane in which an ONeMg WD + MS system may become an eccentric MSP
+ sdB system after AIC. To obtain the birth rate of the eccentric
MSP + sdB systems, I made two binary population synthesis (BPS)
calculations using the rapid binary evolution code developed by
\citet{HUR00,HUR02}. I assume that if the initial orbital period,
$P_{\rm orb}^{\rm i}$, and the initial secondary mass, $M_{\rm
2}^{\rm i}$, of an ONeMg WD + MS system are located in the
appropriate region in the ($\log P^{\rm i}-M_{\rm 2}^{\rm i}$)
plane for the eccentric MSP + sdB systems at the onset of Roche
lobe overflow (RLOF), an eccentric MSP + sdB system can be
produced. I followed the evolution of $10^{\rm 8}$ sample
binaries, where the primordial binary samples are generated in a
Monte Carlo way. The assumptions and the input parameters for the
Monte Carlo simulations are the same to that in \citet{MENGXC17},
and I do not show the repetitious details here. As I will show
in the next section, the common-envelope (CE) ejection efficiency,
$\alpha_{\rm CE}$, is a very important parameter to affect the
birth rate of the eccentric MSP + sdB systems. Following
\citet{MENGXC17}, I take $\alpha_{\rm CE}=1.0$ or $\alpha_{\rm
CE}=3.0$ (see \citealt{MENGXC17} for details)

\begin{figure}
\centerline{\includegraphics[angle=270,scale=.42]{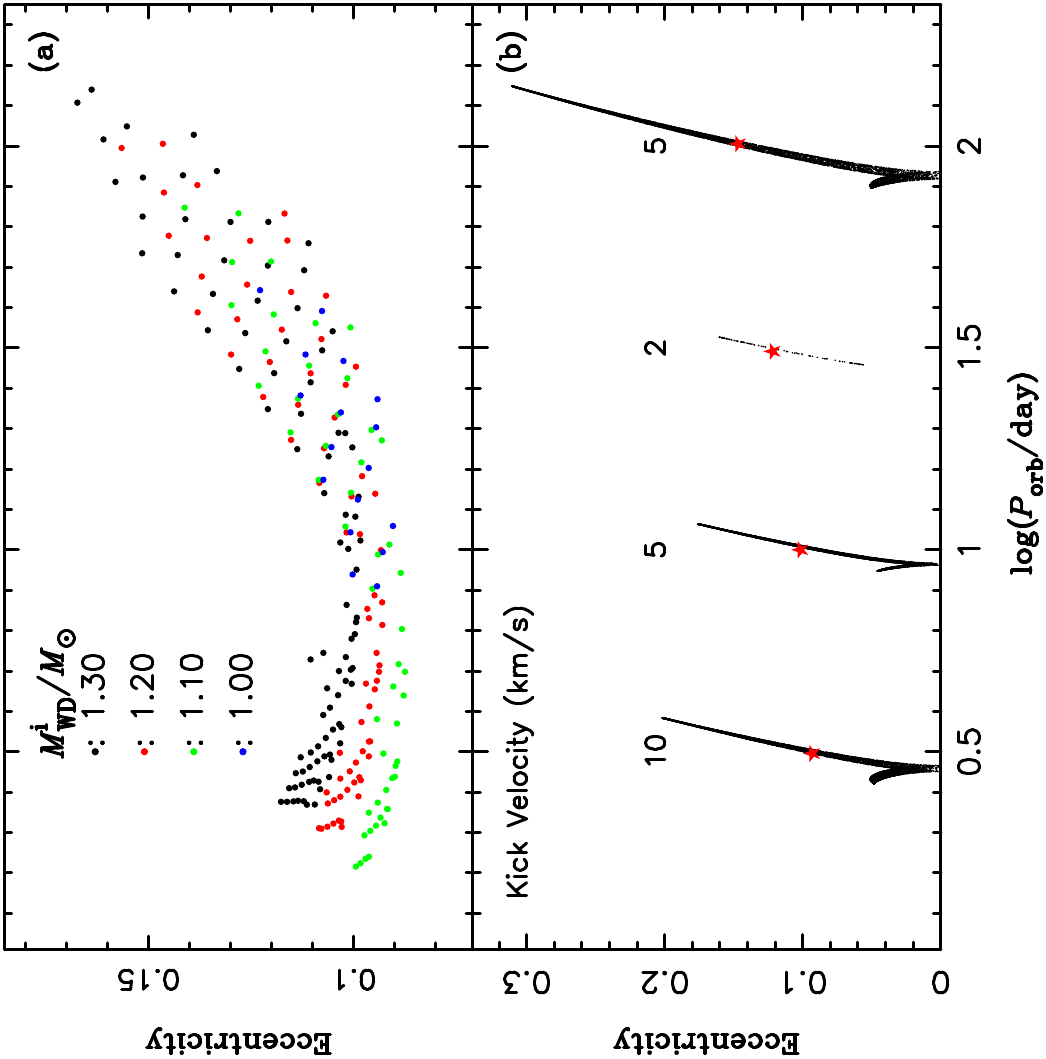}}
\caption{Panel (a): Eccentricities vs orbital periods of the
systems from the RD-AIC scenario for different initial WD masses,
where a symmetric collapse is assumed. Panel (b): Distributions of
the eccentricities and orbital periods of the post-AIC systems
from the pre-AIC systems of [$M_{\rm WD}/M_{\odot},\,M_{\rm
2}/M_{\odot}\,\log(P/{\rm day})$]=(1.5063, 0.4958, 0.4128),
(1.6829, 0.53`5, 0.9115), (1.9070, 0.4046, 1.3820) and (2.3358,
0.4101, 1.8727), respectively. The stars indicate the systems
without kick, while the others show the distributions of the
eccentricities and orbital periods of the systems with different
small kick velocities in unit of km/s as shown by the numbers, where the direction of the
kick velocity is generated by a Monte-Carlo way.}\label{eporb2}
\end{figure}

\section{RESULTS}\label{sect:3}
\subsection{Properties of the eccentric MSP + sdB systems}\label{sect:3.1}
Measuring the mass of a massive NS is a very important way to
constrain the EoS of NSs, and a significant progress has been made
in this field (\citealt{DEMOREST10}; \citealt{CROMARTIE20}). In
Figure~\ref{mnsm2}, I show the masses of MSPs and sdB stars in eccentric
binaries from the RD-AIC scenario for different initial WD masses.
As shown in the figure, the MSPs have a mass between 1.25
$M_{\odot}$ and 2.2 $M_{\odot}$. The maximum MSP mass here is much
larger than that in \citet{FREIRE14} just because I assume that
the pre-AIC WD may be differentially rotating. The sdB stars have
a mass from 0.38 $M_{\odot}$ to 0.65 $M_{\odot}$ as standard
binary model predicted (\citealt{HANZW03}). In the ($M_{\rm NS}$ - $M_{\rm 2}$)  plane, the systems are clearly divided into two branches, which are derived from  the different evolutionary stages of the companions and differences in mass-transfer rates at the onset of RLOF for initial WD + MS systems (see detailed explanation in \citealt{MENGXC17})

In Figure~\ref{eporb2}, I show the distribution of the
eccentricities and the orbital periods of the post-AIC systems.
For a symmetric-collapse assumption (panel a), the RD-AIC scenario
leads to a very narrow range of post-AIC eccentricities:
0.09-0.17, which is slightly larger than that of the eccentric MSP
+ He WD systems from the RD-AIC scenario because of a more massive
pre-AIC WD (\citealt{FREIRE14}). However, the post-AIC orbital
periods here range from $\sim1.6$ days to $\sim150$ days which is
much larger than that for eccentric MSP + He WD systems. The
orbital period of a post-AIC system mainly depends on its pre-AIC
binary evolution and then on the initial binary parameters of a WD
+ MS system (see also \citealt{MENGXC21}).

The eccentricity and orbital period of a post-AIC system is quite
sensitive to a kick to the MSP during the AIC. I use the formulae
in \citet{HILLS83} to check the potential dynamical effect to four different systems by
assuming different small kick velocities, where the direction of
the kick velocity is generated by a Monte-Carlo way. The results
are shown in the panel (b) of Figure~\ref{eporb2}. Similar to that
in \citet{FREIRE14}, the eccentricity and the orbital period of a
given system distribute in a $\surd$-shape region in the
($e-\log P_{\rm orb}$) plane for a relatively larger kick
velocity, while in a narrow strip for a relatively smaller kick
velocity. The panel (b) shows that for a system with an
asymmetrical AIC, the post-AIC eccentricity may be as large as
0.3, even though the kick velocity is as small as 5 ${\rm km\,{\rm
s^{\rm -1}}}$. In addition, as expected in the formulae of
\citet{HILLS83}, for a given value of the kick velocity, the
longer the pre-AIC orbital period, the larger the maximum of  and
then the range of the post-AIC eccentricity.

\begin{figure}
\centerline{\includegraphics[angle=270,scale=.40]{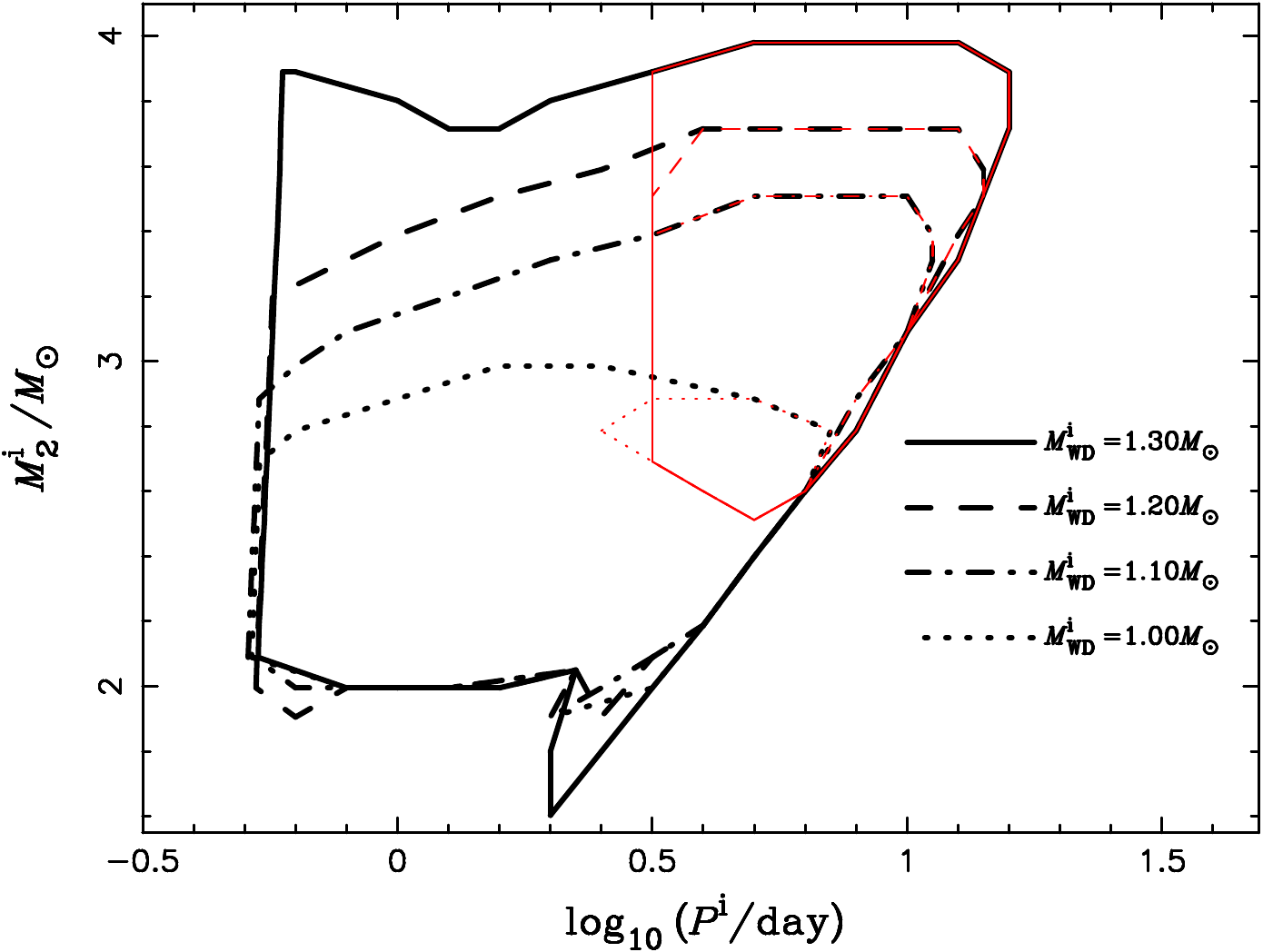}}
\caption{The parameter spaces in ($M_{\rm 2}^{\rm i}-\log P^{\rm
i}$) plane for different initial WD masses, in which a ONeMg WD +
MS system may lead to an eccentric NS + sdB system via RD-AIC
scenario (red lines). The black lines are those for SNe Ia from
\citet{MENGXC17,MENGXC18}.}\label{gaic}
\end{figure}

\begin{figure}
\begin{minipage}[t]{0.5\textwidth}
    \centering
\includegraphics[angle=270,scale=.37]{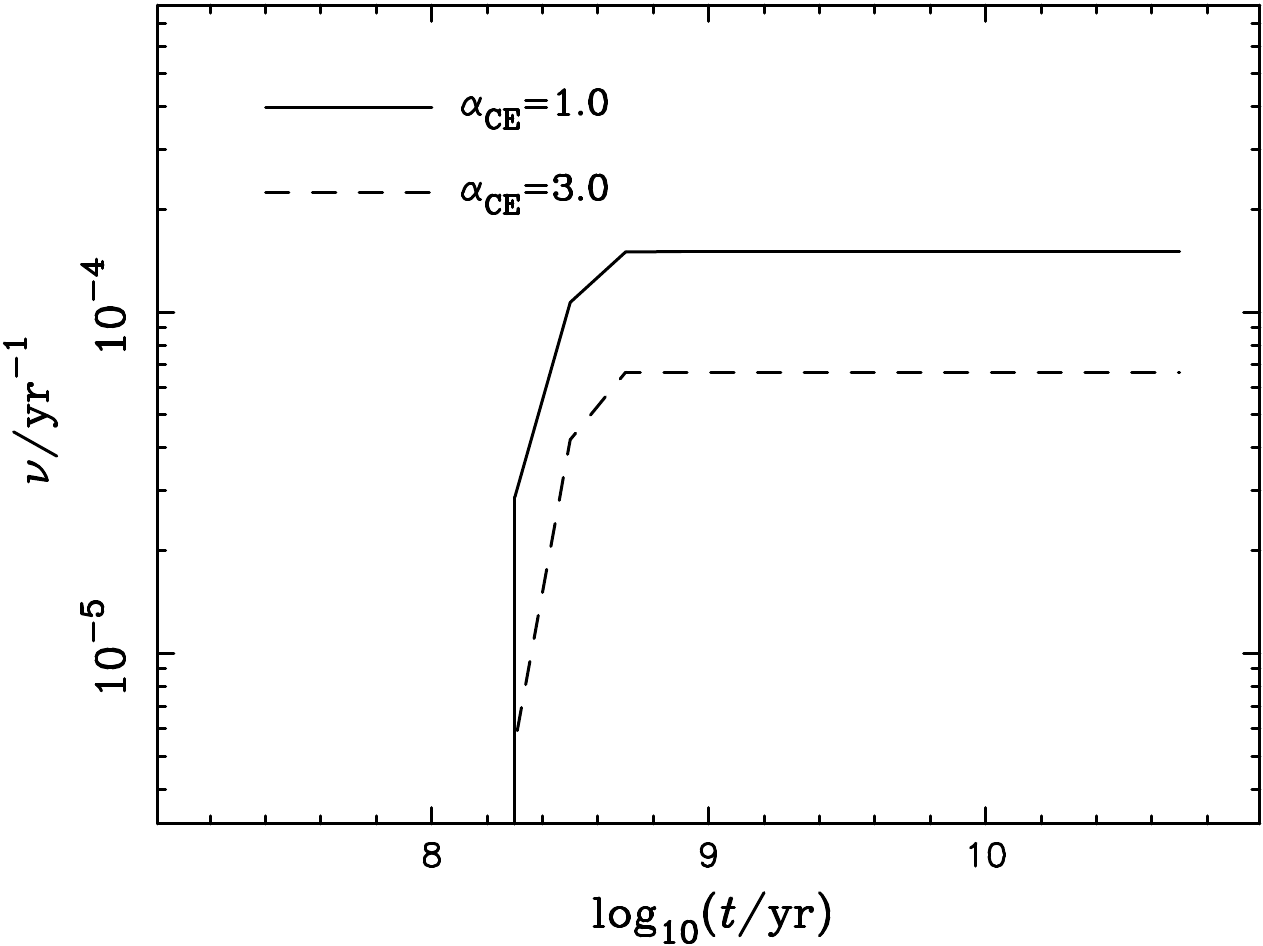}
\end{minipage}
\begin{minipage}[t]{0.5\textwidth}
    \centering
\includegraphics[angle=270,scale=.37]{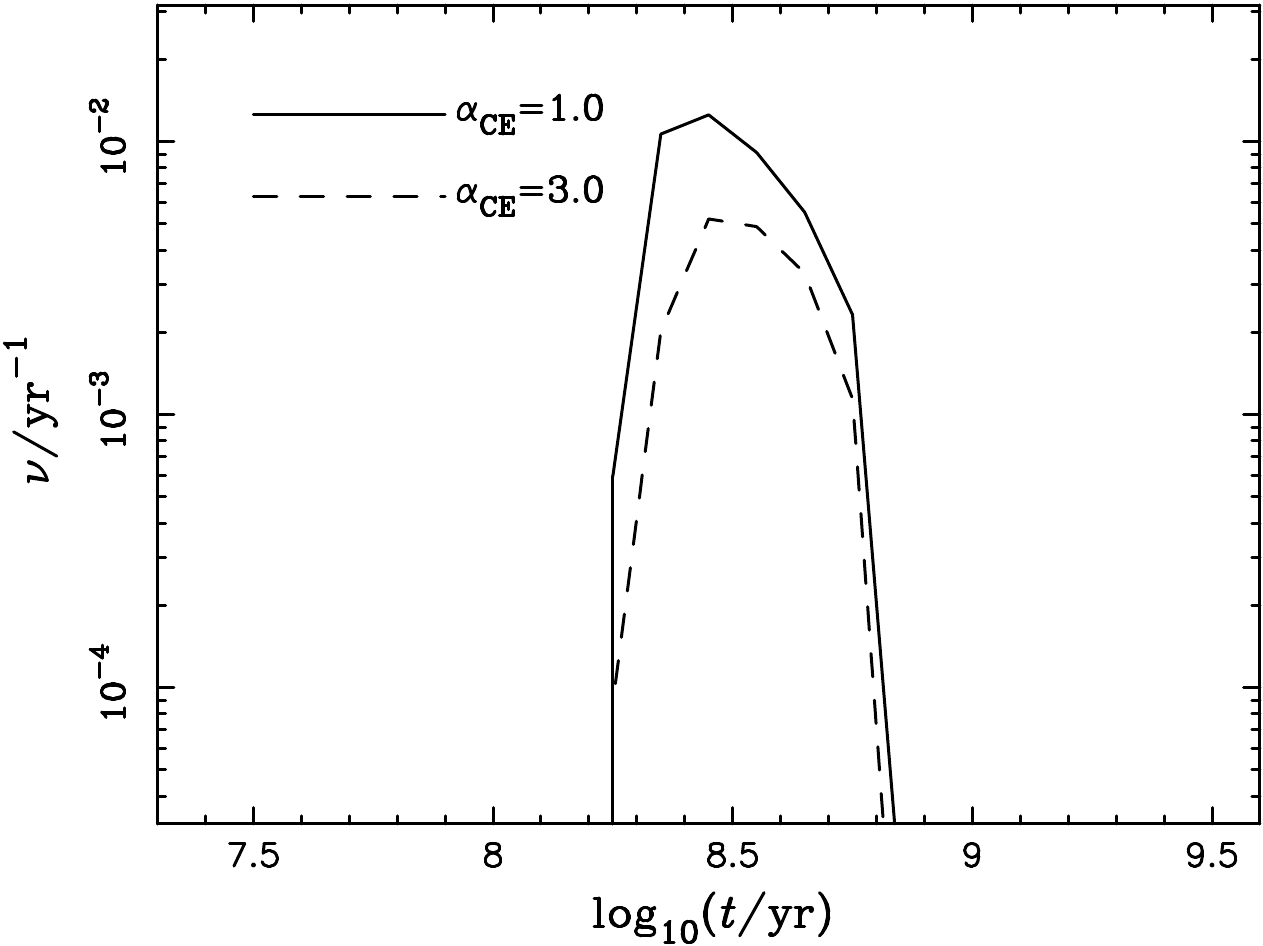}
\end{minipage}
\caption{The evolution of the birth rates, $\nu$, of the eccentric
MSP + sdB systems from the RD-AIC scenario for two values of
$\alpha_{\rm CE}$ and a constant star formation rate of SFR =
$5\,M_{\odot}\,{\rm yr}^{\rm -1}$ (top panel), or a single
starburst of $10^{\rm 11}$ $M_{\odot}$ (bottom panel).}\label{sfr}
\end{figure}

According to the model grid calculations in \citet{MENGXC19a}, in
Figure~\ref{gaic} I show the parameter spaces in ($M_{\rm 2}^{\rm
i}-\log P^{\rm i}$) plane, in which a ONeMg WD + MS system is assumed to lead
to an eccentric MSP + sdB system via the RD-AIC scenario.
Comparing with the parameter spaces for SNe Ia, the parameter
spaces leading to the eccentric MSP + sdB systems locate at the
right-upper region, i.e. their initial orbital periods are longer
than 3 days and initial companion is more massive than 2.5
$M_{\odot}$, which indicates that the eccentric MSP + sdB systems from the RD-AIC scenario belong to relatively young population.

\subsection{Birth rate of eccentric MSP + sdB systems}\label{sect:3.2}
Based on the parameter spaces, I carried out two BPS calculation
to obtain the birth rate of the eccentric MSP + sdB systems. In
the Figure~\ref{sfr}, I show the evolution of the birth rates,
$\nu$, of the eccentric MSP + sdB systems from the RD-AIC scenario
for a constant Galactic star formation rate of SFR = $5\,M_{\odot}\,{\rm
yr}^{\rm -1}$ (top panel) and a single starburst of $10^{\rm 11}$
$M_{\odot}$ (bottom panel). The Galactic birth rate of these
systems is then $(0.67-1.5)\times10^{\rm -4}~{\rm yr^{\rm -1}}$, which
is significantly smaller than the total birth rate of AIC events
(\citealt{WANGB18}). Then, the number of the eccentric MSP + sdB systems may be calculated by $\nu\times\tau_{\rm MSP}$, where $\tau_{\rm MSP}$ is the lifetime of the eccentric MSP + sdB systems. Assuming an optimistic lifetime of $10^{\rm
8}$ yr for the MSPs with strong magnetic fields and a typical life
of $10^{\rm 8}$ yr for a sdB star (\citealt{IVANOVA08}), the
number of the eccentric MSP + sdB systems in the Galaxy is then
6700 - 15000, which is similar to that of NS + sdB systems from
core collapse supernovae (\citealt{WUY20}). The delay time for the
eccentric MSP + sdB systems is between 160 Myr and 630 Myr and the
peak of the birth rate for a single starburst locates at $\sim300$
Myr (bottom panel). So, these systems belong to relatively young
population as their progenitor systems indicate. Assuming the lifetime of $10^{\rm 8}$ yr for sdB stars
and MSPs with strong magnetic fields, there are (6-15) such
systems in a cluster of $10^{\rm 6}$ $M_{\odot}$ with an age of
$\sim300$ Myr. This number is much smaller than that of total AIC
events in a cluster (\citealt{IVANOVA08}). Considering that the
dynamical interaction is very efficient in a cluster and then the
eccentric MSP + sdB systems with a long orbital period are likely
to be disrupted and to possibly form isolated MSPs, such a number estimation could even be too
optimistic. I address this issue in Section~\ref{sect:4} again.

\begin{figure}
\begin{minipage}[t]{0.5\textwidth}
    \centering
\includegraphics[angle=270,scale=.37]{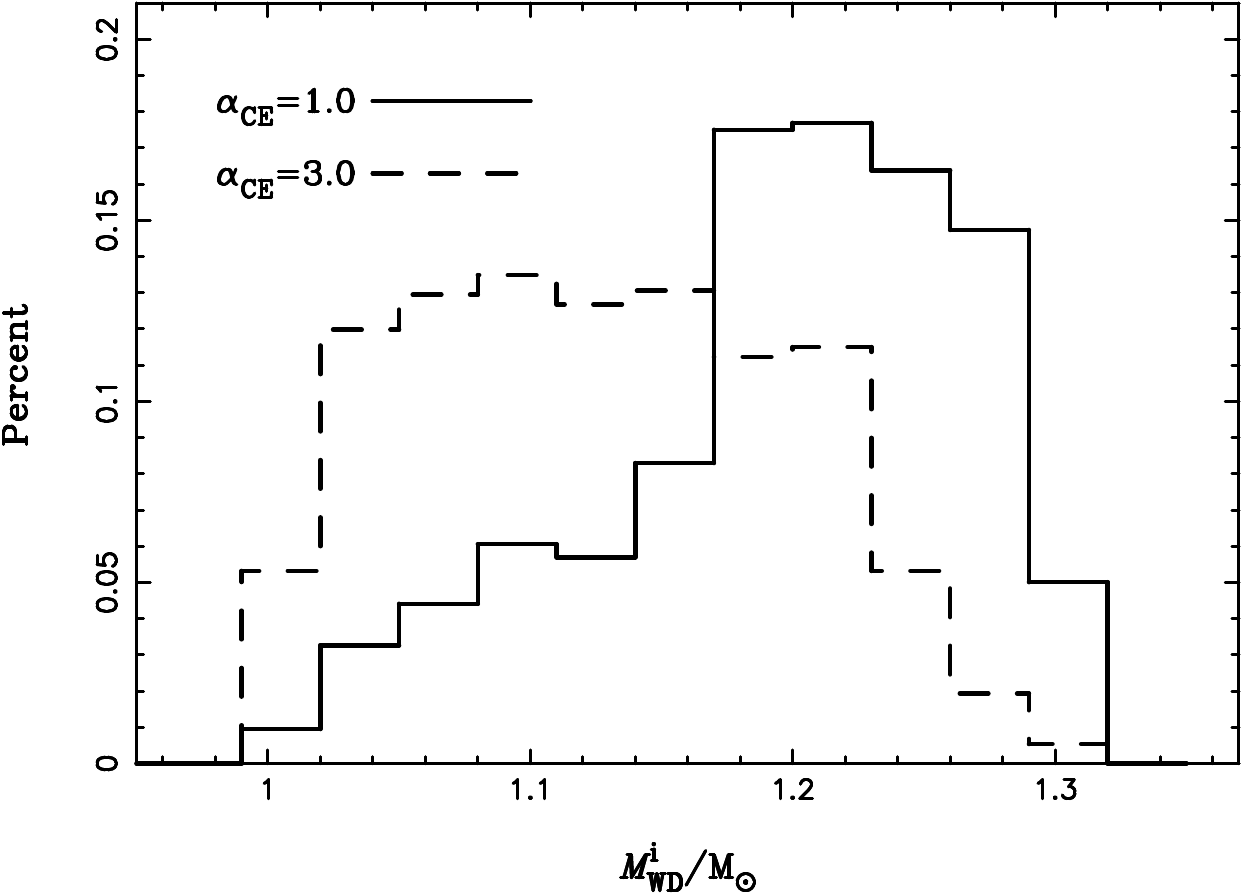}
\end{minipage}
\begin{minipage}[t]{0.5\textwidth}
    \centering
\includegraphics[angle=270,scale=.37]{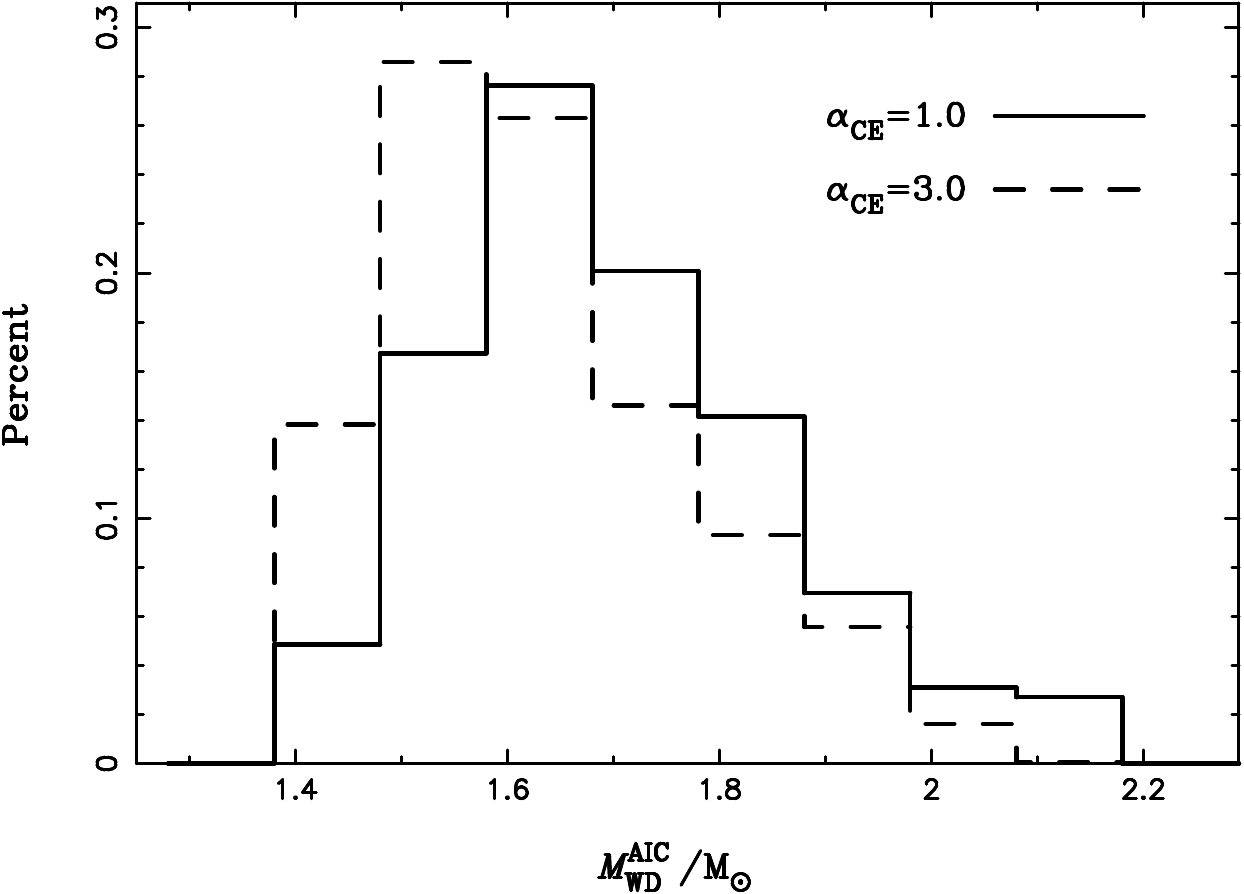}
\end{minipage}
\caption{The distribution of the initial WD mass (top panel) and the WD mass before AIC (bottom panel) for two values of
$\alpha_{\rm CE}$.}\label{mwddis}
\end{figure}

\subsection{Distribution of WDs }\label{sect:3.3}
The main different assumptions between this paper and \citet{FREIRE14} is that the WDs before AIC may be differential rotating, rather than just rigid rotating, and then the WD before AIC may be more massive than 1.48 $M_{\odot}$. However, there might be various efficient angular momentum transfer mechanisms, which means that different rotation patterns would be possible. Generally speaking, the WDs are rigid rotating before AIC if their masses are smaller than 1.48 $M_{\odot}$. Therefore, the mass distribution of white dwarfs prior to AIC may reveal the importance of rigidly rotating white dwarfs in the formation of eccentric MSP + sdB systems. In Figure~\ref{mwddis}, I show the distribution of initial WD mass and the WD mass before AIC for different $\alpha_{\rm CE}$. It is clearly shown in the top panel of Figure~\ref{mwddis} that a smaller $\alpha_{\rm CE}$ leads to the peak of the initial WD mass distribution to move to a higher value. This is derived from a fact that all the systems experience a CE phase before they become the ONeMg WD + MS systems. A system with a less massive WD is more likely to merge for a smaller $\alpha_{\rm CE}$ value after a common envelope phase, i.e. as a result, a system with a more massive WD is more likely to survive from the common envelope evolution and then becomes a potential progenitor system of an  eccentric MSP + sdB system.

For the distribution of the WD mass before AIC, three properties are worthy to be noticed in the bottom panel of Figure~\ref{mwddis} . First, only about 5\% to 14\% WDs have a mass less than 1.48 $M_{\odot}$, which indicates that the rigid rotation is not a dominant case for the AIC channel. This is mainly from the fact that an ONeMg WD generally has a large initial mass and then is easy to exceed the limit of the rigid rotation after a mass transfer occurs between the WD and its massive companion (more massive than 2.5 $M_{\odot}$ and see also Figure~\ref{gaic}). Second, the peak of the distribution moves to a lower value for a higher $\alpha_{\rm CE}$, which is directly correlated with the distribution of the initial WD mass (see the top panel of Figure~\ref{mwddis}). Third, about 50\% to 70\% WD have a mass less massive than about 1.68 $M_{\odot}$ and only about 2\% to 7\% WDs have a mass more massive than 2 $M_{\odot}$, although in principle, a 2.6 $M_{\odot}$ WD is possible before AIC. This means that most of the MSPs in the eccentric MSP + sdB systems have a mass less massive than $\sim1.5$ $M_{\odot}$, which provides another information to be verified by a large sample of the eccentric MSP + sdB systems in the future.

\begin{figure}
\centerline{\includegraphics[angle=270,scale=.40]{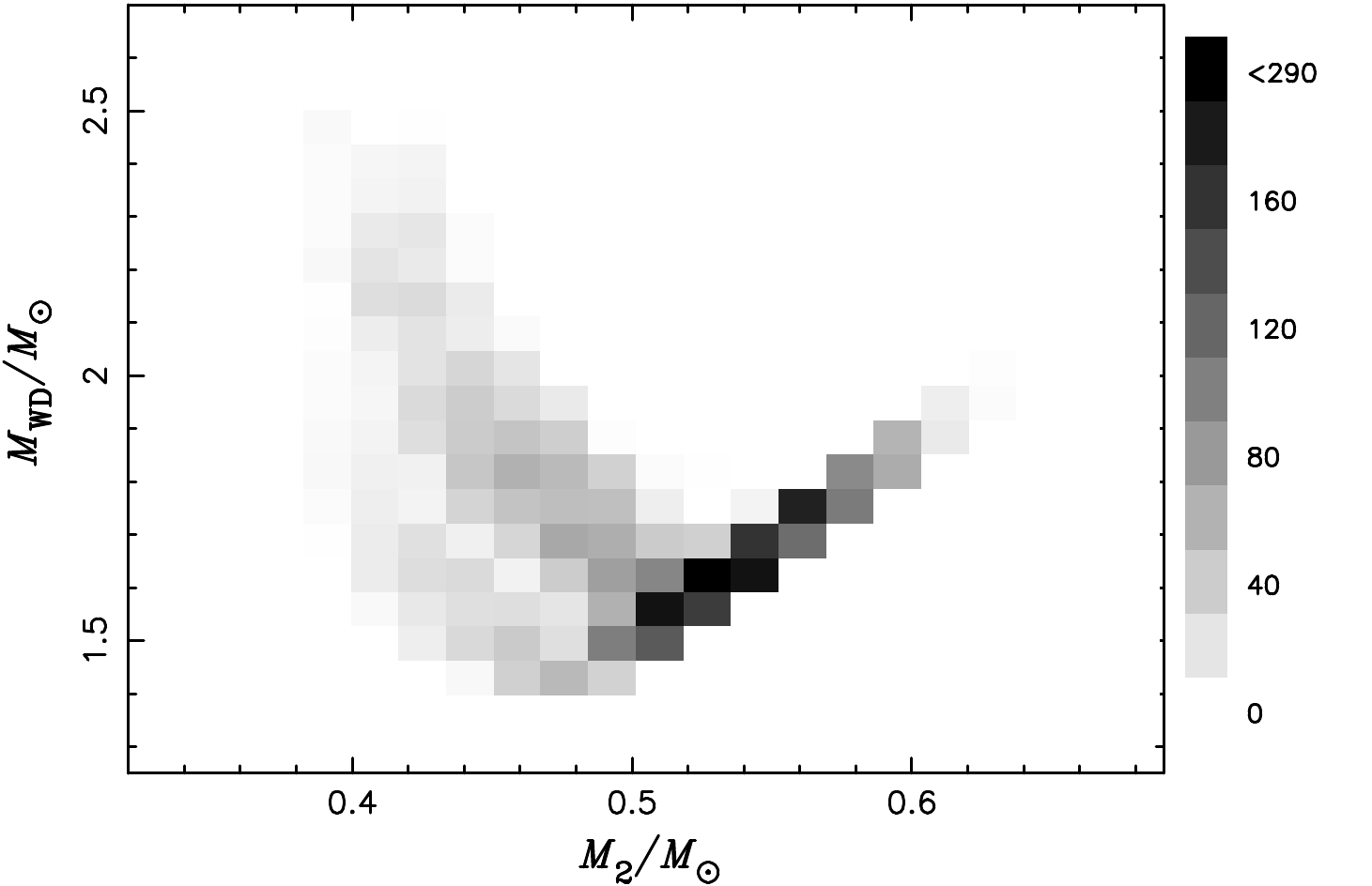}}
\caption{The distributions of WD and sdB masses at the moment of AIC in the $M_{\rm WD}$-$M_{\rm 2}$ plane,
where $\alpha_{\rm CE}=1.0$. The basic properties for the case of $\alpha_{\rm CE}=3.0$ is similar.}\label{m2mwd}
\end{figure}

Figure~\ref{m2mwd} presents the distributions of WD and its companion masses at the moment of AIC in the $M_{\rm WD}$-$M_{\rm 2}$ plane. Here, I only present the case of $\alpha_{\rm CE}=1.0$ and the case of $\alpha_{\rm CE}=3.0$ is similar to that. According to Figure~\ref{m2mwd},  readers may also obtain the distributions of the MSP and sdB masses in eccentric MSP + sdB systems in a $M_{\rm NS}$-$M_{\rm 2}$ plane, as shown in Figure~\ref{mnsm2}. Similar to Figure~\ref{mnsm2},  the systems are also clearly divided into two branches in the $M_{\rm WD}$-$M_{\rm 2}$ plane, i.e. no system locates at the region around $M_{\rm WD}\sim2.1$ $M_{\odot}$ and $M_{\rm 2}\sim0.52$ $M_{\odot}$. For the WDs of less than 1.5 $M_{\odot}$, the masses of sdB stars focus on $\sim0.5$ $M_{\odot}$, while for the WDs of higher than 2.0 $M_{\odot}$, the sdB stars have a mass less than 0.48  $M_{\odot}$. The two-branch distribution here provides a way to examine the predictions in this paper by a large sample in the future.

\section{DISCUSSIONS AND CONCLUSIONS}\label{sect:4}

In this study, inspired by the suggestion of \citet{FREIRE14} that the RD-AIC scenario may account for the formation of eccentric MSP + He WD systems, I estimate that there could be 6700-15000 eccentric MSP +
sdB systems in the Galaxy. Their orbital parameters are also predicted; for instance, under the assumption of symmetric collapse, their eccentricities lie between 0.09 and 0.17 (see Figures~\ref{mnsm2} and \ref{eporb2}), with most MSPs having masses below 1.5 $M_{\odot}$. The formation of such systems requires mass transfer to initiate during the HG phase in an initial  ONeMg WD + MS bianry. The delay time for their emergence is on the order of several hundred Myr, peaking at $\sim300$ Myr. Therefore, these systems are expected to be found in relatively young environments such as the Galactic thin disk. In particular, if the associated AIC involves only a low kick velocity, the resulting systems are likely to possess low peculiar velocities and follow disk-like orbits in the Galaxy. A future detection of eccentric MSP + sdB binaries would thus offer additional indirect evidence in support of the AIC process.

\subsection{Eccentric MSP + He WD systems }\label{sect:4.1}

\begin{figure}
\centerline{\includegraphics[angle=270,scale=.40]{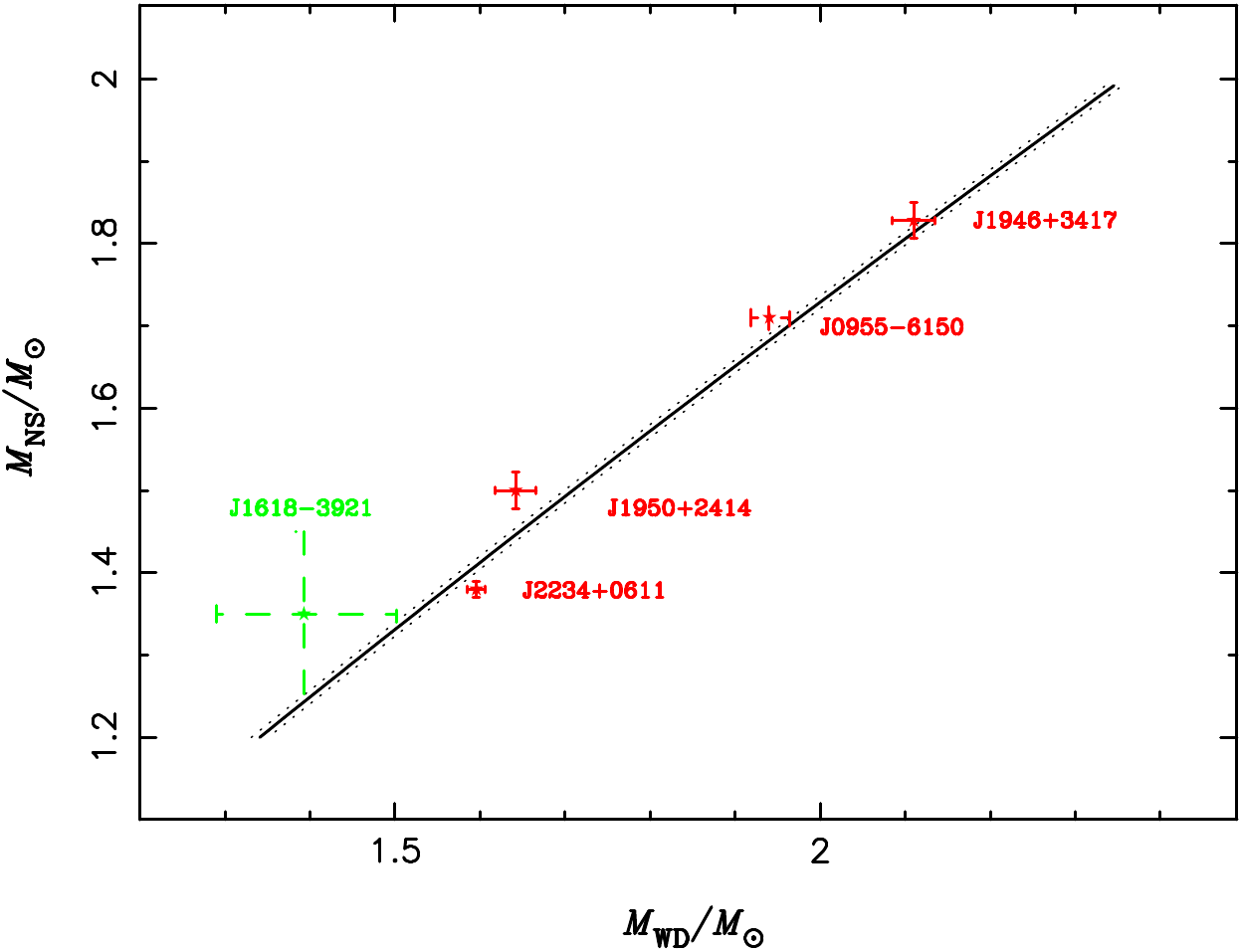}}
\caption{Pre-AIC WD mass vs. post-AIC NS mass with an assumption of
differential rotation, symmetric collapse and ejection of baryonic
material of 0.02 $M_{\odot}$ (solid line), or 0.01 $M_{\odot}$ and
0.03 $M_{\odot}$ (dotted line), respectively. The stars present observed
results, where the symmetric collapse is assumed. Red stars show
the systems with precise measurement of NS mass and eccentricity,
while the green one shows the results where the NSs
are assumed to have a mass of 1.25-1.45 $M_{\odot}$ with a median
of 1.35 $M_{\odot}$ and the companions are assumed to be He WDs.
The observational dada are from \citet{BARR17}, \citet{OCTAU18},
\citet{STOVALL19}, \citet{ZHUWW19} and
\citet{SERYLAK22}.}\label{mnsmwd}
\end{figure}

To account for the observed properties of the eccentric MSP + He WD systems, \citet{FREIRE14} adopt a WD delay time longer than $10^{\rm 9}$ yrs, while a pseudo-spin-down timescale of $10^{\rm 7}$ yrs is assumed in this work. The extended delay time in \citet{FREIRE14} is motivated by two considerations. First, prior to AIC, the red giant companion is assumed to evolve into a white dwarf, terminating mass transfer. Consequently, after AIC, the resulting eccentric MSP + He WD system may retain its orbital eccentricity. Second, a longer delay allows the super-Chandrasekhar WD sufficient time to cool before undergoing RD-AIC. Furthermore, \citet{FREIRE14} assume rigidly rotating super-Chandrasekhar ONeMg WDs, which restricts their masses to below 1.48 $M_{\odot}$ (\citealt{YOON04}). In contrast, our model incorporates differentially rotating WDs, permitting masses exceeding 2 $M_{\odot}$, as  illustrated in Figures~\ref{mwddis} and \ref{m2mwd}. If our assumptions prove viable for eccentric MSP + sdB systems, they may, in principle, also offer a plausible explanation for the observed eccentric MSP + He WD systems.

The assumed spin-down timescale of $10^{\rm 7}$ yrs is consistent with, and at least does not conflict with, the two constraints mentioned above. First, a delay time of $\sim10^{\rm 7}$ yrs  is certainly sufficient to allow a WD of $\sim 0.3$ $M_{\odot}$ to enter the cooling branch of WD before AIC occurs (\citealt{CHENLQ17}; \citealt{CHENXF17}). Second, although the timescale to undergo RD-AIC remains highly uncertain, i.e in a range of $10^{\rm 5}$ yrs to $10^{\rm 9}$ yrs, depending on the redistribution of angular momentum within super-Chandrasekhar ONeMg WDs (\citealt{YOON04,YOON05}), it is still plausible for an AIC to occur in a super-Chandrasekhar ONeMg WD in $\sim10^{\rm 7}$ yrs, particularly in systems hosting WDs as massive as  2.4 $M_{\odot}$ (\citealt{HACHISU12}; also \citealt{MENGXC13}).

The mass of a NS in an eccentric binary system can be determined through measurements of relativistic Shapiro delay or the rate of advance of periastron (\citealt{DEMOREST10}; \citealt{OZEL16}; \citealt{CROMARTIE20}).  While some properties of MSP systems are well explained by the RD-AIC scenario (\citealt{FREIRE14}; \citealt{WANGD23}), others remain challenging, for instance, some observed pulsar masses substantially exceed the predictions under the rigid-rotation RD-AIC hypothesis proposed by \citet{FREIRE14} (\citealt{BARR17}; \citealt{SERYLAK22}). If eccentric MSP systems originate from a differential-rotation RD-AIC channel, the orbital eccentricity could be used to infer the pre-AIC WD mass of the NS via Equation~(\ref{eq:1}). Consequently, the measurement of the NS mass and the eccentricity may shed light on the formation origin of eccentric MSPs. As an illustration, Figure~\ref{mnsmwd} presents the relation between pre-AIC WD mass and post-AIC NS mass under the assumption of symmetric collapse, alongside observational data for eccentric MSP + He WD systems (also assuming symmetric collapse). These systems align closely with the theoretical relation predicted by the RD-AIC scenario, suggesting that provided differentially rotating ONeMg WDs exist, such systems could indeed originate from RD-AIC. Similarly, future discoveries of eccentric MSP + sdB systems could further help to clarify the origin of eccentric MSP + He WD binaries.  Another noteworthy feature in Figure~\ref{mnsmwd} is that three of the five observed eccentric MSP + He WD systems contain MSPs with masses below 1.5 $M_{\odot}$, consistent with the prediction shown in Figure~\ref{mwddis}. This highlights the need for future detailed BPS studies that incorporate differential rotation, rather than relying solely on rigid-rotation assumptions (e.g. \citealt{WANGD23}). Additionally, detailed binary evolution calculations confirm that super-Chandrasekhar WDs exceeding 2 $M_{\odot}$ indeed form from WD + red giant systems when differential rotation is taken into account (\citealt{CHENLQ17}). Finally, the slight deviation of PSR J1950+2414 from the theoretical relation in Figure~\ref{mnsmwd} may indicate the presence of a small kick imparted to the MSP during the AIC process.

\subsection{Uncertainties and potential applications for eccentric MSP + sdB systems}\label{sect:4.2}
Compared to the circular MSP + He WD binaries, systems with eccentric orbits are exceptionally rare, accounting for only about 2\%. In addition to the RD-AIC channel, neutron stars originating from core-collapse supernovae may also form MSP + sdB systems via the standard recycling process (\citealt{WUY18}). Based on BPS simulations in \citet{WUY20}, I estimate that the fraction of eccentric MSP + sdB systems could be as high as 55\% among all MSP + sdB binaries, which is significantly larger than the 2\% observed for MSP + He WD systems. However, the predicted Galactic population of eccentric MSP + sdB systems is highly sensitive to the lifetime of strongly magnetized MSPs, with $10^{\rm 8}$ yr likely representing a conservative upper limit. For instance, magnetars can have characteristic ages as short as $10^{\rm 3}$ yr (\citealt{IGOSHEV21}), implying that the expected number of eccentric MSP + sdB systems arising from the RD-AIC scenario may be less than one. Furthermore, the spin-down timescale preceding AIC also plays a critical role in determining the abundance of such systems. In this work, I adopt a pseudo-spin-down timescale of $\tau_{\rm sp}=10^{\rm 7}$ yr, motivated by observational constraints from type Ia supernovae. In principle, however, $\tau_{\rm sp}$ prior to an AIC event could be considerably longer than $10^{\rm 7}$ yr, given the fundamental physical differences between AIC and type Ia explosions. Should $\tau_{\rm sp}$ substantially exceed $10^{\rm 8}$ yr, the sdB companion would evolve into a CO WD before AIC occurs. In that case, the RD-AIC channel would not produce eccentric MSP + sdB systems, but rather eccentric MSP/NS + CO WD binaries. Consequently, the numbers predicted here for eccentric MSP + sdB systems should be regarded as a very conservative upper limit. Even if all post-AIC MSPs formed via RD-AIC exhibited lifetimes typical of normal MSPs, the population of eccentric MSP + sdB systems would still be roughly an order of magnitude smaller than that of normal MSP binaries (\citealt{BHATTACHARYYA21}). Finally, future detections of eccentric MSP + sdB systems may help clarify whether the observed prevalence of very low magnetic fields among MSPs is intrinsic or a selection effect resulting from their longer lifetimes (\citealt{FREIRE13}).

A comparison between the predicted upper limit on the total number of eccentric MSP + sdB binaries in the Galaxy and the currently observed population of such systems could impose a strong constraint on either the lifetime of post-AIC MSPs or the spin-down timescale prior to AIC. Currently, there are approximately 550 known regular MSPs\footnote{https://pages.astro.umd.edu/~eferrara/GalacticMSPs.html} and an underlying population of MSPs of about 83000 (\citealt{LEVIN13}; \citealt{LORIMER13}).
Assuming no significant selection effects against their detection, these represent $550/83000\sim0.66$\% of all regular MSPs in the Galaxy. Based on this detection fraction, and considering that the predicted orbital characteristics of eccentric MSP + sdB systems are not too extreme, one would expect to observe between 45 and 100 such systems. Yet, none have been detected.  This non-detection implies that the lifetime of eccentric MSP + sdB systems formed via AIC could be shorter than $10^{\rm 6}$ yr, or alternatively, that the pseudo-spin-down timescale for AIC is close to $10^{\rm 8}$ yr. Consequently, based on the observational constraints, the upper limit of the number of such systems could be 1500.

In such systems, the sdB stars possess masses between 0.38 $M_{\odot}$ and 0.65 $M_{\odot}$, and therefore typically do not undergo a red-giant phase (\citealt{JUSTHAM11b}). Given their relatively long orbital periods, these eccentric MSP + sdB binaries will avoid a stable Roche lobe overflow mass-transfer phase, meaning their orbital eccentricity remains unchanged until they evolve into eccentric NS + CO WD systems. Consequently, the population of eccentric NS + CO WD systems originating from eccentric MSP + sdB progenitors is inherently older than that of the MSP + sdB systems themselves. A future statistical comparison between the observed populations of eccentric MSP + sdB systems and their descendant eccentric NS + CO WD systems could thus provide a meaningful test of the RD-AIC scenario. Regrettably, no systems matching these predictions have yet been identified in the Galactic field.

One question remains to be addressed: why have eccentric MSP + He WD systems been observed, but not eccentric MSP + CO WD systems in Galactic field? Three facts may contribute to this. First, the optimistic estimated number of eccentric MSP + sdB systems, and hence the optimistic number of potential eccentric MSP + CO WD systems originating from the RD–AIC scenario, is at least one order of magnitude smaller than that of eccentric MSP + He WD systems (\citealt{WANGD23}). Second, even under the most optimistic estimates for the population of eccentric MSP + sdB systems, the resulting eccentric MSP + CO WD descendants are expected to be rare, since, as discussed earlier, such MSPs are likely to evolve into normal pulsars or neutron stars in graveyards. Additionally, it is crucial to note that the very conservative upper limit of 55\% for the eccentric fraction among MSP + sdB systems cannot be directly extrapolated to the MSP + CO WD population. This is because the latter forms through multiple channels, and the RD–AIC scenario likely constitutes only a minor fraction of these systems. (\citealt{TAURIS12}; \citealt{CHENWC13}; \citealt{WANGB17}; \citealt{WANGB20}). 

It is widely believed that the MSPs observed in globular clusters (GCs) originate from the AIC channel (\citealt{VANDENHEUVEL10}). However, given the advanced ages of GCs, no eccentric MSP + sdB systems have been detected within them to date. Such systems would eventually evolve into a normal pulsar + CO WD stage while retaining their orbital eccentricity. Notably, systems like J1750-3703A and J1824-2452D may represent this population, as they exhibit moderately eccentric orbits, host a slowly rotating pulsar, and have companion mases and orbital periods consistent with theoretical predictions\footnote{Please see P. Freire's
Web site http://www.naic.edu/~pfreire/GCpsr.html.}.

Moreover, \citet{PODSIADLOWSKI05} proposed that if a NS originates from an electron - capture collapse, measuring its mass could help constrain the EoS of NSs by examining the relationship between the pre - collapse and post - collapse masses. Similarly, a pre - AIC to post - AIC mass relation could also constrain the NS EoS, since the AIC process is itself a form of electron - capture collapse. The
eccentric MSP + sdB systems, if they are indeed from the RD-AIC scenario, offer a promising avenue for such constraints, as both the pre-AIC and post-AIC masses of the NS may be determined with high precision.

Spinning NSs with non-axisymmetric distortions resulting from solid deformation or excited fluid oscillation modes can emit continuous gravitational waves (CGWs) (\citealt{ALFORD15}; \citealt{CHENWC21}). An asymmetric rapid RD-AIC event could produce such distortions, potentially leading to detectable CGW signals in future observing runs of Advanced LIGO and Virgo (\citealt{ABBOT20}). Combined with precise NS mass measurements, the detection of CGWs could offer direct insights into the EoS of NSs (\citealt{GLAMPEDAKIS18}).

Finally, the results presented here rely on the standard common-envelope wind model, in which several parameters remain highly uncertain, for instance, the adopted CE density (\citealt{MENGXC17}). Given the uncertainties associated with the CE density, the lower limit of the orbital period shown in Figure ~\ref{eporb2} could be overestimated by up to an order of magnitude.

\section*{ACKNOWLEDGMENTS}
This work is supported by the National Natural Science Foundation of China (Nos. 12288102, 12333008),  the Strategic Priority Research Program of the Chinese Academy of Sciences (grant Nos. XDB1160303, XDB1160000), and the National Science Foundation of China and National Key R\&D Program of China (No. 2021YFA1600403). X.M. acknowledges support from Yunnan Fundamental Research Projects (NOs. 202401BC070007 and 202201BC070003), International Centre of Supernovae, Yunnan Key Laboratory (No. 202302AN360001), Yunnan Revitalization Talent Support Program - Yunling Scholar Project and the Yunnan Revitalization Talent Support Program - Science \& Technology Champion Project (NO. 202305AB350003),  and the China Manned Space Program with grant No. CMS-CSST-2025-A13.

\section*{Data Availability}

No new data were generated in support of this research, and all
the data used in this article will be shared on request to the
corresponding author.


\begin{thebibliography}{}
 \bibitem[\protect\citeauthoryear{Abbott {\it et al.}}{2020}]{ABBOT20}
Abbott, R., Abbott, T.D., Abraham, S. et al., 2020, ApJL, 902, L21
\bibitem[\protect\citeauthoryear{Alford \& Schwenzer}{2015}]{ALFORD15}
Alford, M.G. \& Schwenzer, K., 2015, MNRAS, 446, 3631
\bibitem[\protect\citeauthoryear{Alpar {\it et al.}}{1982}]{ALPAR82}
Alpar, M.A., Cheng, A.F., Ruderman, M.A., \& Shaham, J., 1982,
Natur, 300, 728
\bibitem[\protect\citeauthoryear{Antoniadis}{2014}]{ANTONIADIS14}
Antoniadis, J., 2014, ApJL, 797, L24
\bibitem[\protect\citeauthoryear{Archibald et al.}{2009}]{ARCHIBALD09}
Archibald, A.M., Stairs, I.H., Ransom, S.M., et al., 2009,
Science, 324, 1411
\bibitem[\protect\citeauthoryear{Barr {\it et al.}}{2017}]{BARR17}
Barr, E.D., Freire, P.C.C., Kramer., M. et al., 2017, MNRAS, 465,
1711
\bibitem[\protect\citeauthoryear{Bhattacharyya \& van den Heuvel}{1991}]{BHATTACHARYA91}
Bhattacharyya, D. \& van den Heuvel, E.P.J., 1991, PhR, 203, 1
\bibitem[\protect\citeauthoryear{Bhattacharyya \& Roy}{2021}]{BHATTACHARYYA21}
Bhattacharyya, B. \& Roy, J., 2021, arXiv: 2104.02294, Preprint of
a chapter of the book 'Millisecond Pulsars', of the Astrophysics
and Space Science Library (ASSL) series edited by Sudip
Bhattacharyya, Alessandro Papitto and Dipankar Bhattacharya
\bibitem[\protect\citeauthoryear{Champion}{2008}]{CHAMPION08}
Champion, D.J., Ransom, S.M., Lazarus, P., et al., 2008, Science,
320, 1309
\bibitem[\protect\citeauthoryear{Chen \& Liu}{2013}]{CHENWC13}
Chen, W.C. \& Liu, W.M., 2013, MNRAS, 432, L75
\bibitem[\protect\citeauthoryear{Chen}{2021}]{CHENWC21}
Chen, W.C., 2021, PhRvD, 103, 103004
\bibitem[\protect\citeauthoryear{Chen {\it et al.}}{2017}]{CHENLQ17}
Chen L., Meng X. C., Han Z. W., 2017, RAA, 17, 8, 83 
\bibitem[\protect\citeauthoryear{Chen {\it et al.}}{2017}]{CHENXF17}
Chen X. F.,  Maxted P. F. L.,  Li J., Han Z. W., 2017, MNRAS, 467, 1874 
\bibitem[\protect\citeauthoryear{Cromartie et al.}{2020}]{CROMARTIE20}
Cromartie, H.T., Fonseca, E., Ransom, S.M. et al., 2020, NatAs, 4,
72
\bibitem[\protect\citeauthoryear{Demorest et al.}{2010}]{DEMOREST10}
Demorest, P.B., Pennucci, T., Ransom, S.M. et al., 2010, Nature,
467, 1081
\bibitem[\protect\citeauthoryear{Di Stefano \& Kilic}{2012}]{DISTEFANO12}
Di Stefano R., \& Kilic M. 2012, ApJ, 759, 56
\bibitem[\protect\citeauthoryear{Freire {\it et al.}}{2011}]{FREIRE11}
Freire, P.C.C., Bassa, C.G., Wex, N., et al., 2011, MNRAS, 412,
2763
\bibitem[\protect\citeauthoryear{Freire}{2013}]{FREIRE13}
Freire, P.C.C., 2013, IAUS, 291, 243
\bibitem[\protect\citeauthoryear{Freire \& Tauris}{2014}]{FREIRE14}
Freire, P.C.C. \& Tauris, T.M., 2014, MNRAS, 438, L86
\bibitem[\protect\citeauthoryear{Glampedakis \& Gualtieri}{2018}]{GLAMPEDAKIS18}
Glampedakis, K. \& Gualtieri, L., 2018, in The Physics and
Astrophysics of Neutron Stars, Astrophysics and Space Science
Library, Vol. 457, ed. L. Rezzolla et al. (Dordrecht: Springer),
673
\bibitem[\protect\citeauthoryear{Geier {\it et al.}}{2011}]{GEIER11}
Geier, S., Heber, U., Tillich, A., et al. 2011, in AIP Conf. Ser., eds. S. Schuh, H.
Drechsel, \& U. Heber, 1331, 163
\bibitem[\protect\citeauthoryear{Geier {\it et al.}}{2015}]{GEIER15}
Geier S. et al., 2015, Science, 347, 1126
\bibitem[\protect\citeauthoryear{Geier {\it et al.}}{2017}]{GEIER17}
Geier, S., \O {}stensen, R. H., Nemeth, P., et al. 2017, A\&A, 600, A50
\bibitem[\protect\citeauthoryear{Ginzburg \& Chiang}{2022}]{GINZBURG21}
Ginzburg, S. \& Chiang, E., 2022, MNRAS, 509, L1
\bibitem[\protect\citeauthoryear{Ginzburg \& Chiang}{2020}]{GOTBERG20}
G\"{o}tberg Y., Korol V., Lamberts A. et al., 2020, ApJ, 904, 14
\bibitem[\protect\citeauthoryear{Guo et al.}{2024}]{GUOYL24}
Guo Y., Wang B., Li X., 2024, MNRAS, 527, 7394
\bibitem[\protect\citeauthoryear{Guo et al.}{2025}]{GUOYL25}
Guo Y., Wang B., Li X., Liu D., Tang W., 2025, arXiv: 2505.16299
\bibitem[\protect\citeauthoryear{Hachisu {\it et al.}}{2012}]{HACHISU12}
Hachisu I., Kato, M., Saio H.,  Nomoto K., 2012, ApJ, 744, 69
\bibitem[\protect\citeauthoryear{Han \& Li}{2021}]{HANQ21}
Han, Q. \& Li, X.D., 2021, ApJ, 909, 161
\bibitem[\protect\citeauthoryear{Han {\it et al.}}{2007}]{HANZW07}
Han Z., Podsiadlowski P., \& Lynas-Gray A. E. 2007, MNRAS, 380, 1098
\bibitem[\protect\citeauthoryear{Han et al.}{2003}]{HANZW03}
Han, Z., Podsiadlowski, Ph., Maxted, P. F. L., Marsh, T. R., 2003,
MNRAS, 341, 669
\bibitem[\protect\citeauthoryear{He {\it et al.}}{2025}]{HERJ25}
He R., Meng X., Lei Z., Yan H., Lan S., 2025, A\&A,  693, A121
\bibitem[\protect\citeauthoryear{Heber}{2009}]{HEBER09}
Heber, U., 2009, ARA\&A, 47, 211
\bibitem[\protect\citeauthoryear{Heber}{2016}]{HEBER16}
Heber, U., 2016, PASP, 128, 082001
\bibitem[\protect\citeauthoryear{Hills}{1983}]{HILLS83}
Hills J. G., 1983, ApJ, 267, 322
\bibitem[\protect\citeauthoryear{Hurley {\it et al.}}{2000}]{HUR00}
Hurley, J.R., Pols, O.R., Tout, C.A., 2000, MNRAS, 315, 543
\bibitem[\protect\citeauthoryear{Hurley et al.}{2002}]{HUR02}
Hurley, J.R., Tout, C.A., Pols, O.R., 2002, MNRAS, 329, 897
\bibitem[\protect\citeauthoryear{Ivanova {\it et al.}}{2008}]{IVANOVA08}
Ivanova, N., Heinke, C.O., Rasio, F.,A., Belczynski, K., Fregeau,
J.M., 2008, MNRAS, 386, 553
\bibitem[\protect\citeauthoryear{Igoshev {\it et al.}}{2021}]{IGOSHEV21}
Igoshev, A.P., Popov, S.B., Hollerbach, R., 2021, Universe, 7, 351
\bibitem[\protect\citeauthoryear{Jiang et al.}{2015}]{JIANGL15}
Jiang, L., Li, X.D., Dey, J., et al., 2015, ApJ, 807, 41
\bibitem[\protect\citeauthoryear{Justham}{2011}]{JUSTHAM11}
Justham S., 2011, ApJL, 730, L34
\bibitem[\protect\citeauthoryear{Justham {\it et al.}}{2011}]{JUSTHAM11b}
Justham, S., Podsiadlowski, Ph., Han, Z., 2011b, MNRAS, 410, 984
\bibitem[\protect\citeauthoryear{Kremer {\it et al.}}{2024}]{KREMER24}
Kremer, K., Ye, C.S., Heinke, C.O. et al., 2024, ApJ, 977, L42
\bibitem[\protect\citeauthoryear{Lattimer \& Yahil}{1989}]{LATTIMER89}
Lattimer, J.M. \& Yahil, A., 1989, ApJ, 340, 426
\bibitem[\protect\citeauthoryear{Levin {\it et al.}}{2013}]{LEVIN13}
Levin, L.,  Bailes, M.,  Barsdell, B. R. et al., 2013, MNRAS, 434, 1387
\bibitem[\protect\citeauthoryear{Lorimer}{2013}]{LORIMER13}
Lorimer, D.R., 2013, IAUS, 291, 237
\bibitem[\protect\citeauthoryear{Manchester {\it et al.}}{2005}]{MANCHESTER05}
Manchester, R.N., Hobbs, G.B., Teoh, A., et al. 2005, AJ, 129,
1993
\bibitem[\protect\citeauthoryear{Meng \& Podsiadlowski}{2013}]{MENGXC13}
Meng, X. \& Ph. Podsiadlowski, 2013, ApJL, 778, L35
\bibitem[\protect\citeauthoryear{Meng \& Podsiadlowski}{2017}]{MENGXC17}
Meng, X. \& Podsiadlowski, Ph. 2017, MNRAS, 469, 4763
\bibitem[\protect\citeauthoryear{Meng \& Podsiadlowski}{2018}]{MENGXC18}
Meng, X. \& Podsiadlowski, Ph. 2018, ApJ, 861, 127
\bibitem[\protect\citeauthoryear{Meng \& Li}{2019}]{MENGXC19a}
Meng, X. \& Li, J., 2019, MNRAS, 482, 5651
\bibitem[\protect\citeauthoryear{Meng \& Luo}{2021}]{MENGXC21}
Meng, X. \& Luo, Y., 2021, MNRAS, 507, 4603
\bibitem[\protect\citeauthoryear{Moehler {\it et al.}}{1990}]{MOEHLER90}
Moehler S., Richtler T., de Boer K. S., Dettmar R. J., \& Heber U. 1990,
A\&AS, 86, 53
\bibitem[\protect\citeauthoryear{Octau {\it et al.}}{2018}]{OCTAU18}
Octau, F., Cognard, I., Guillemot, L. et al., 2018, A\&A, 612, A87
\bibitem[\protect\citeauthoryear{\"{O}zel \& Freire}{2016}]{OZEL16}
\"{O}zel, F. \& Freire, P., 2016, ARA\&A, 54, 401
\bibitem[\protect\citeauthoryear{Oostrum {\it et al.}}{2020}]{OOSTRUM20}
Oostrum, L. C., van Leeuwen, J., Maan, Y., Coenen, T.,
Ishwara-Chandra, C. H., 2020, MNRAS, 492, 4825
\bibitem[\protect\citeauthoryear{Phinney \& Kulkarni}{1994}]{PHINNEY94}
Phinney, E. S., \& Kulkarni, S. R., 1994, ARA\&A, 32, 591
\bibitem[\protect\citeauthoryear{Podsiadlowski {\it et al.}}{2005}]{PODSIADLOWSKI05}
Podsiadlowski, Ph., Dewi, J.D.M., Lesaffre, P., Miller, J.C.,
Newton, W.G., Stone, J.R., 2005, MNRAS, 361, 1243
\bibitem[\protect\citeauthoryear{Prada Moroni \& Straniero}{2009}]{PRADA09}
Prada Moroni P.G.,  Straniero O., 2009, A\&A, 507, 1575
\bibitem[\protect\citeauthoryear{Radhakrishnan \& Srinivasan}{1982}]{RADHAKRISHNAN82}
Radhakrishnan, V. \& Srinivasan, G., 1982, Curr. Sci., 51, 1096
\bibitem[\protect\citeauthoryear{Serylak {\it et al.}}{2022}]{SERYLAK22}
Serylak, M., Venkatraman Krishnan, V., Freire, P.C.C. et al.,
2022, A\&A, 665, A53
\bibitem[\protect\citeauthoryear{Stovall{\it et al.}}{2019}]{STOVALL19}
Stovall, K., Freire, P.C.C., Antoniadis, J. et al., 2019, ApJ,
870, 74
\bibitem[\protect\citeauthoryear{Tang {\it et al.}}{2023}]{TANGWS23}
Tang, Wen-Shi, Gao, Shi-Jie, Li, Xiang-Dong, 2023, MNRAS, 519, 2951
\bibitem[\protect\citeauthoryear{Tauris {\it et al.}}{2012}]{TAURIS12}
Tauris T. M., Langer N.,  Kramer M., 2012, MNRAS, 425, 1601
\bibitem[\protect\citeauthoryear{Tauris {\it et al.}}{2013}]{TAURIS13}
Tauris, T.M., Sanyal, D.,  Yoon, S.C., Langer, N., 2013, A\&A,
558, A39
\bibitem[\protect\citeauthoryear{van den Heuvel}{2010}]{VANDENHEUVEL10}
van den Heuvel, E.P.J., 2010, NewAR, 54, 140
\bibitem[\protect\citeauthoryear{Wang {\it et al.}}{2017}]{WANGB17}
Wang, B., Podsiadlowski, Ph., \& Han, Z. 2017, MNRAS, 472,
1593
\bibitem[\protect\citeauthoryear{Wang}{2018}]{WANGB18}
Wang, B., 2018, MNRAS, 481, 439
\bibitem[\protect\citeauthoryear{Wang \& Liu}{2020}]{WANGB20}
Wang, B. \& Liu, D.D., 2020, RA\&A, 20,135
\bibitem[\protect\citeauthoryear{Wang \& Gong}{2023}]{WANGD23}
Wang, D. \& Gong, B. P., 2023, MNRAS, 526, 5021
\bibitem[\protect\citeauthoryear{Wijnands \& van der Klis}{1998}]{WIJNANDS98}
Wijnands, R. \& van der Klis, M., 1998, Nature, 394, 344
\bibitem[\protect\citeauthoryear{Wickramasinghe \& Ferrario}{2005}]{WF05}
Wickramasinghe, D.T. \& Ferrario, L., 2005, MNRAS, 356, 1576
\bibitem[\protect\citeauthoryear{Wu {\it et al.}}{2018}]{WUY18}
Wu Y., Chen X., Li Z., Han Z., 2018, A\&A, 618, A14
\bibitem[\protect\citeauthoryear{Wu {\it et al.}}{2020}]{WUY20}
Wu Y., Chen X., Chen H., Li Z., Han Z., 2020, A\&A, 634, A126
\bibitem[\protect\citeauthoryear{Yang {\it et al.}}{2025}]{YANGYZL25}
Yang, Z.L., Han, J.L., Zhou, D.J. et al., 2025, Science, 388, 859
\bibitem[\protect\citeauthoryear{Yoon \& Langer}{2004}]{YOON04}
Yoon, S.-C. \& Langer, N., 2004, A\&A, 419, 623
\bibitem[\protect\citeauthoryear{Yoon \& Langer}{2005}]{YOON05}
Yoon, S.-C. \& Langer, N., 2005, A\&A, 435, 967
\bibitem[\protect\citeauthoryear{Zhu {\it et al.}}{2019}]{ZHUWW19}
Zhu, W.W., Freire, P.C.C., Knispel, B. et al., 2019, ApJ, 881, 165
  
   
   
\end{thebibliography}
\end{document}